# Isotopically enriched epitaxial CaWO$_4$ thin films for Er$^{3+}$ spin-photon quantum interfaces

Hanlin Tang[1], Kidae Shin[2], Ashwin K. Boddeti[3,4], Sebastian P. Horvath[3], Adam Turflinger[3], Joseph Alexander[3], Jeffrey A. Dhas[5], Zihua Zhu[6], Shuhang Pan[2], Jeff D. Thompson[3], Yingge Du[5], Frederick J. Walker[1], and Charles H. Ahn[*][1,2,7]

1) Department of Applied Physics, Yale University, New Haven, CT 06520, USA.

2) Department of Physics, Yale University, New Haven, CT 06520, USA.

3) Department of Electrical and Computer Engineering, Princeton University, Princeton, NJ 08544

4) Center for Integrated Nanotechnologies, Sandia National Laboratories, Albuquerque, NM, 87185, USA.

5) Physical Sciences Division, Pacific North National Laboratory, Richland, Washington 99354, USA.

6) Environmental Molecular Sciences Laboratory, Pacific North National Laboratory, Richland, Washington 99354, USA.

7) Department of Materials Science, Yale University, New Haven, CT 06520, USA.

Rare earth ion (REI)-doped oxide thin films are attractive for the application of quantum interconnects due to their stable optical levels and scalability[1–3]. Among them, Er$^{3+}$ doped CaWO$_4$ is promising because it possesses narrow optical linewidth transitions and a long spin coherence time[4–6]. The electron spin coherence is limited at high temperatures by paramagnetic impurities and by the presence of the 14.3% $^{183}$W nuclear spin. To further increase the spin coherence time at millikelvin temperatures, where the paramagnetic impurities are frozen out, our approach is to synthesize chemically and isotopically purified thin films as a host material. We first grow non-isotopically enriched Er$^{3+}$ doped CaWO$_4$ thin films, which exhibit a 214(13) MHz photoluminescence (PL) inhomogeneous linewidth, indicating the thin film has high crystalline quality. We then grow isotopically enriched CaWO$_4$ thin films using an isotopically purified $^{186}$WO$_3$ source. Time of flight secondary ion mass spectrometry (ToF-SIMS) was used to measure the relative concentration of W isotopes. $^{183}$W, the only W isotope that has a net nuclear spin and is the major cause of spin decoherence, was at a relative abundance of 1.2%, a factor of 10 lower than natural abundance. We also observed PL emission from single ions after integrating nano-photonic devices with the thin film. These results establish isotopically engineered CaWO$_4$ thin films as a promising platform for future studies of nuclear-spin-limited coherence and for scalable rare-earth-ion-based quantum nanophotonic devices.

[*]charles.ahn@yale.edu

## Introduction

Dielectric oxides doped with rare-earth ions (REIs) are promising for quantum interconnect (QuICs) applications because of their highly coherent optical transitions.[7–11] Also, the 4f-4f optical transitions and Kramers doublets of REIs in solid-state host systems provide a natural approach to realize spin-photon entanglement, a key component in QuICs[12,13], when the REI is used as a single-ion optical emitter that can be addressed using nanophotonic cavities[14]. High-performance QuICs require long-lived spin states, a narrow optical homogeneous linewidth, and optical transitions in the low-loss telecom band[7]. Among REIs, Er$^{3+}$ is particularly attractive because of its $^4I_{15/2} \rightarrow {}^4I_{13/2}$ optical transition (1.53 $\mu m$) in the telecom C-band. Examples of Er$^{3+}$ in various single crystal oxide hosts include Er$^{3+}$: TiO$_2$ [15–17], Er$^{3+}$ : YVO$_4$[18], Er$^{3+}$: CaWO$_4$[4,19,20], Er$^{3+}$: and CaMoO$_4$[21]. Among them, CaWO$_4$ shows considerable potential as an Er$^{3+}$ host for a number of reasons. First, the absence of a permanent electric field at the Er$^{3+}$ doping site means that the optical homogeneous linewidth broadening and spectral diffusion due to the first-order Stark effect are suppressed, enabling the generation of indistinguishable photons[4] and radiative lifetime-limited coherence[22]. Second, as a result of low nuclear spin concentration in natural abundance CaWO$_4$ (14.3% $^{183}$W), a long spin coherence time $T_2$ up to 23ms has been observed in Er$^{3+}$:CaWO$_4$ at millikelvin temperatures[5].

An improvement of the spin coherence can be achieved by eliminating the nuclear spin bath using isotopically pure elements[5,23–26]. In the case of Er$^{3+}$ doped in CaWO$_4$ as a single photon source, the use of isotopically pure bulk materials is inefficient because only the top few nanometers of the surface are coupled to nanophotonic devices. We show that an efficient use of isotopically pure source material can be achieved by thin film growth using Molecular Beam Epitaxy (MBE). Approaches to grow nuclear spin-free oxide thin films using natural abundance source materials, including CeO$_4$/Si[27–29], cannot achieve the coherence time predicted by previous cluster-correlation expansion (CCE) calculations, which only consider the decoherence from the nuclear spin bath[30]. This limitation is likely due to the spectral diffusion induced by magnetic dipolar interactions from the $Er^{3+}$ ensemble, charged defects, impurities, and strain-induced decoherence. Isotopically purified homoepitaxial $CaWO_4$ thin films, on the other hand, provide a nearly nuclear spin-free system, excluding the interface strain effect from the substrate. To date, the impact of nuclear-spin purification on spin and optical coherence has not been experimentally explored in $CaWO_4$ thin films, motivating the present study.

Here, we use MBE to grow homoepitaxial $CaWO_4$ and measure a 214 (13) MHz inhomogeneous optical linewidth in a PL spectrum, indicating that the thin films we synthesize possess high crystalline quality. We apply this approach to grow isotopically enriched $Ca^{186}WO_4$ thin films with 2 grams of $^{186}WO_3$ source material. We also find that defects that result from a fluctuating $WO_3$ flux can be eliminated by a post-annealing method, as confirmed by High-resolution phase-contrast Transmission Electron Microscopy (HRTEM). Isotopic analysis using ToF-SIMS measurements shows that the nuclear spin concentration in the isotopically enriched thin film has been reduced by a factor of 10 compared with the substrates. Finally, we realize optical addressing of single $Er^{3+}$ ions in MBE-grown $CaWO_4$ thin films, demonstrating compatibility of the thin-film platform with single-emitter nanophotonic integration.

## Method

**MBE thin film growth**

Thin films are grown on commercial high-purity (99.9999%) $CaWO_4$ substrates (Surface Net GmbH). Non-isotopically enriched thin films with 30 nm thickness are grown using metallic Ca pellets (99.99%, Sigma Aldrich) and $WO_3$ (14.3% $^{183}WO_3$, Sigma Aldrich) pellet source material with natural abundance distribution of isotopes in an oxide molecular beam epitaxy chamber with $2\times10^{-9}$ Torr base pressure, at a substrate temperature of 600°C, and an oxygen partial pressure of $4 \times 10^{-6}$ Torr. The $WO_3$ and CaO fluxes are measured using a quartz-crystal microbalance (QCM) prior to the growth in $O_2$ and adjusted to have a 1:1 CaO to WO$_3$ flux ratio as measured by the QCM. Isotopically enriched thin films with the same thickness are grown with isotopically enriched $^{186}$WO$_3$ powder (not pellets) with 0.22% $^{183}$WO$_3$, supplied by the National Isotope Development Center. An excess CaO flux of 0%-25%, as measured by the QCM, is used to determine the optimal growth conditions, which may compensate for different tooling factors for the Ca and WO$_3$ sources. Growth is monitored in-situ with a reflection high-energy electron diffraction (RHEED) system at 10 keV.

**ToF-SIMS**

ToF-SIMS depth profiling is performed using a ToF-SIMS V (ION-TOF, GmbH, Münster, Germany) mass spectrometer. For negative ion spectral collection, a 25 keV Bi$^+$ beam (~1.1 $p$A, 100 μm x 100 μm) is employed. The sputtering beam is a 1 keV Cs$^+$ beam (~20 nA, 300 μm x 300 μm). Prior to data acquisition, the analysis beam scanning area is aligned to the center of the sputtering crater. During the depth profile collection, a non-interlaced mode strategy (sputter time of 2.0 seconds and pause time of 0.5 seconds) is used. Charge compensation with an electron flood gun (~1 μA) is also used.

# Results and Discussion

**Photoluminescence and X-ray diffraction on natural abundance MBE-grown thin films**

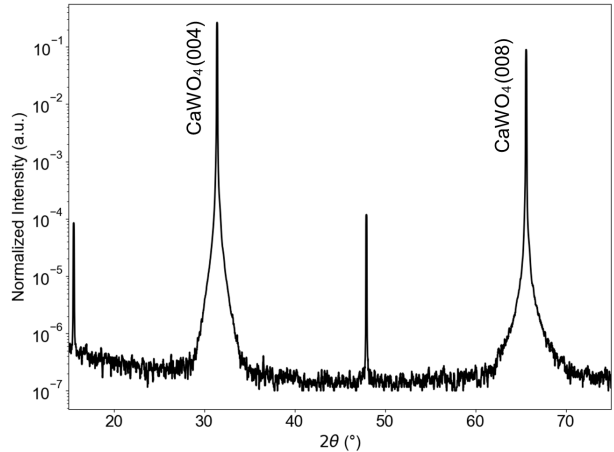

Fig. 1. **Single-phase growth of CaWO₄ thin films.** Narrow peaks shown at diffraction forbidden angles are due to dynamic scattering.

We first grow non-isotopically enriched CaWO$_4$ thin films [31]. A typical X-ray $\theta - 2\theta$ scan in Fig. 1 shows a pure CaWO$_4$ thin film phase. Peaks appearing at diffraction-forbidden positions are due to dynamical diffraction effects, as determined by the observation of a dependence of the intensity on a varying azimuthal angle $\phi$. The lack of finite thickness oscillations near the substrate Bragg reflections indicates that the thin film preserves the same crystal structure and mass density as the substrate. To quantify the crystalline quality of the thin film, we conducted ensemble photoluminescence measurements on Er$^{3+}$ implanted ($1\times10^{12}$ cm$^{-2}$ fluence @ 35keV) into the thin film CaWO$_4$ host. The inhomogeneous linewidth of the Z$_1$-Y$_1$ optical transition is measured to be 214(13) MHz (Fig. 2). Compared to a 730MHz inhomogeneous linewidth in bulk CaWO$_4$,[4] the narrow inhomogeneous linewidth we measure is consistent with high crystalline quality in the thin film, as defects and strain variations would otherwise broaden the optical transition.

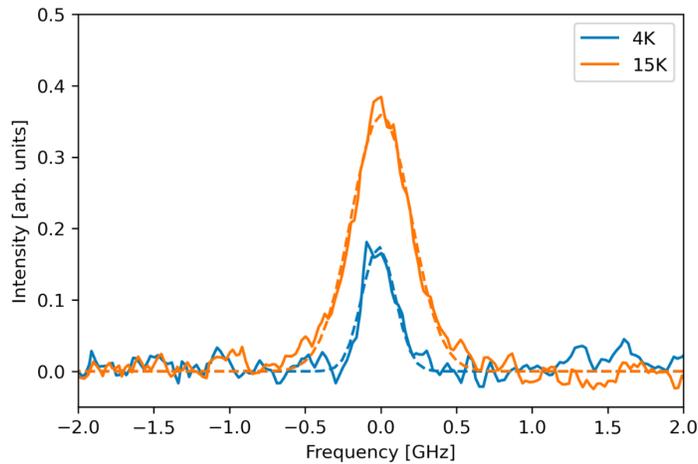

Fig 2. **Narrow PL inhomogeneous linewidth of MBE-grown $Er^{3+}$: $CaWO_4$ thin films at 4K and 15K.** The measured inhomogeneous linewidths are 214 (13) MHz and 370 (8) MHz at 4K and 15K, respectively.

**X-ray diffraction and AFM morphologies of isotopically enriched MBE-grown thin films**

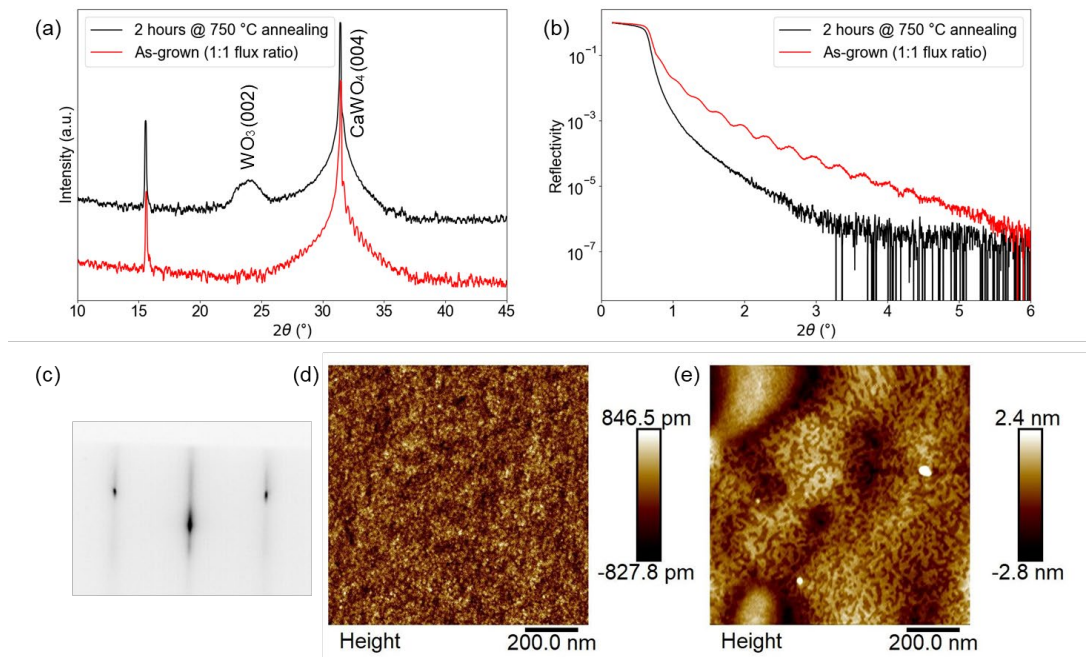

Fig. 3. **The effects of annealing and off-stoichiometric growth.** (a) X-ray diffraction (XRD) $\theta - 2\theta$ scan of as-grown and annealed CaWO$_4$ thin films. A WO$_3$ (002) peak appears after annealing in the tube furnace for 2 hours at 750°C. (b) X-ray reflectivity of as-grown and annealed CaWO$_4$ thin films. (c) RHEED of CaWO$_4$ thin film after growth. (d) AFM surface topography after growth with roughness $R_q = 0.3 nm$. (e) AFM surface topography after annealing.

We then grow isotopically enriched CaWO$_4$ thin films with isotopically enriched WO$_3$ powder. We find that the WO$_3$ flux evaporated from a powder source is less stable compared with a pellet source. A flux measurement over hours using a QCM (see supplement Fig. S1) exhibits WO$_3$ flux fluctuations, which may be caused by a time-varying evaporation surface area. In this case, thin films grown with a 1:1 CaO to WO$_3$ QCM flux ratio show off-stoichiometric features: Kiessig fringes in X-ray reflectivity measurements (Fig. 3(b)) indicate the thin film density is different from the substrate. Moreover, X-ray diffraction scans of the as-grown thin films show Laue oscillations, which are a signature of off-stoichiometric thin films or a layer with defects that modify the lattice parameter and/or density of the film relative to the substrate. Despite oscillations observed in both XRR and XRD scans, the as-grown thin film keeps the CaWO$_4$ scheelite structure, as single-phase and orientation CaWO$_4$ diffraction features are observed in the XRD scan, as well as in the RHEED image (Fig. 3(c)). To improve the crystal quality and eliminate defects, we perform post-annealing of the as-grown films in pure oxygen in a tube furnace for 2 hours at 750°C. X-ray diffraction scans of the annealed thin films show an extra WO$_3$ (002) peak and a CaWO$_4$ (004) peak without Laue oscillations (Fig. 3 (a)), implying the thin film is a two phase mixture of extra WO$_3$ and single phase CaWO$_4$. To investigate the annealing effect on the surface morphology of the thin film, we measure Atomic Force

Microscopy (AFM) before and after thermal annealing. The as-grown thin film has a smooth surface with roughness $R_q = 0.30$nm. After annealing, an increase in roughness is observed in the AFM scan (Fig. 3(e)), likely due to the $WO_3$ inclusions embedded in the thin film.

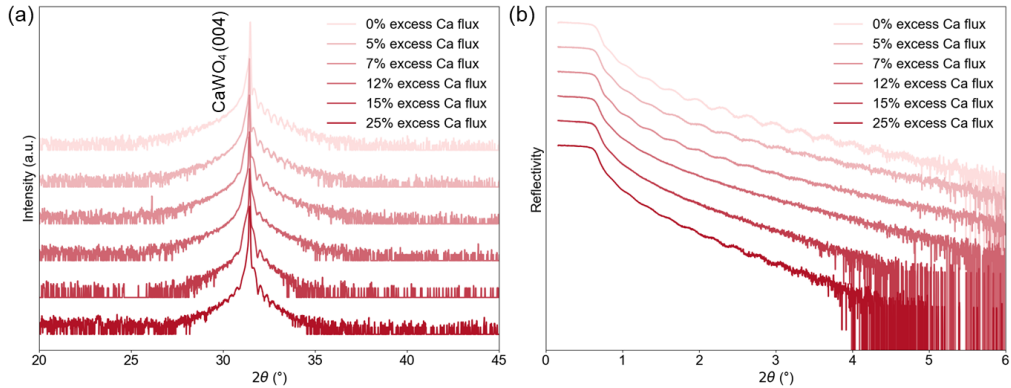

Fig. 4. **Optimizing growth flux based on X-ray characterization.** (a) X-ray diffraction $\theta - 2\theta$ scan around the $CaWO_4$ (004) peak. (b) X-ray reflectivity measurement.

To compensate for the observed off-stoichiometric growth of the thin film, we targeted $CaWO_4$ compositions with 0%-25% excess CaO as measured by the QCM. XRD and XRR measurements of thin films grown using different measured flux ratios are shown in Fig. 4. Among them, the thin film grown using 12% excess CaO has a smooth X-ray reflectivity curve, suggesting the average density and stoichiometry of the thin film match those of the substrate. Small Laue oscillations around the $CaWO_4$ substrate (004) peak may originate from local crystal distortions due to deviations of the thin film lattice from the bulk. As shown in Fig. 5, the surface morphology of the thin film grown using 12% CaO flux is similar to that grown with a 1:1 flux ratio. Both of them have smooth surfaces ($R_q \sim$ 0.3 nm) and sharp RHEED patterns.

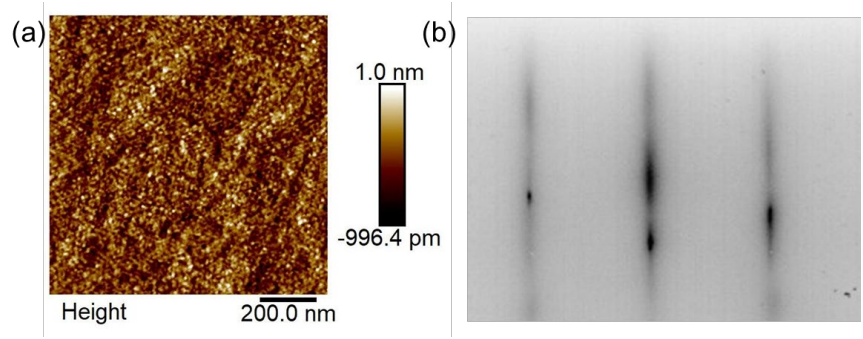

Fig. 5. **Surface characterization of CaWO$_4$ thin films grown with 12% excess CaO flux.** (a) AFM surface morphology of the thin film with roughness $R_q = 0.35$ nm. (b) RHEED pattern after growth, clear and sharp diffraction spots indicate the thin film has high surface crystalline quality.

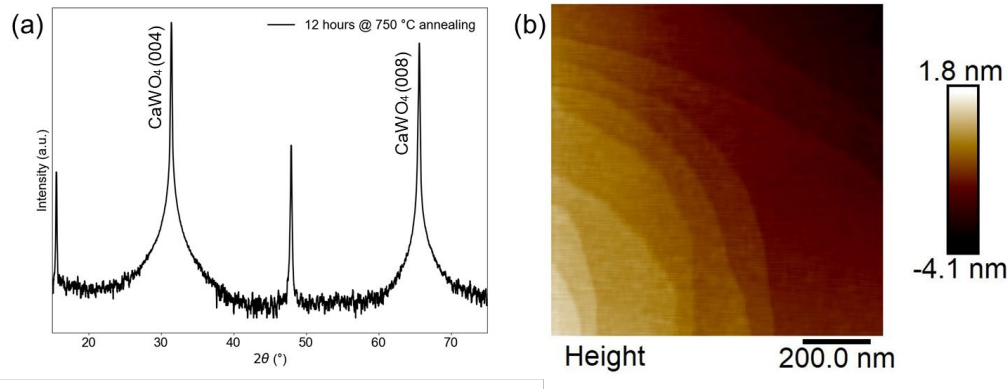

Fig. 6. **Effect of annealing stoichiometric CaWO$_4$ thin films.** (a) X-ray diffraction $\theta - 2\theta$ scan shows no Laue oscillations. (b) $1.0 \times 1.0 \ \mu m^2$ AFM scan of annealed CaWO$_4$ thin film shows smooth step-terrace surface morphology.

After growth, the thin films are annealed in pure O$_2$ at 750°C for 12 hours. XRD of annealed thin films shows a smooth peak and a single CaWO$_4$ phase (Fig. 6(a)), verifying the thin film is stoichiometric and has the same crystal structure as the substrate. It is likely that, during the annealing process, excess Ca atoms in W-deficient regions, resulting from flux instability during the growth, migrate to Ca-deficient regions and restore the stoichiometry. An AFM scan after annealing (Fig. 6(b)) shows clear atomic steps and terraces, with roughness $R_q < 0.07$ nm on one of the terraces, demonstrating significant diffusion at the anneal temperature.

**Transmission Electron Microscopy (TEM)**

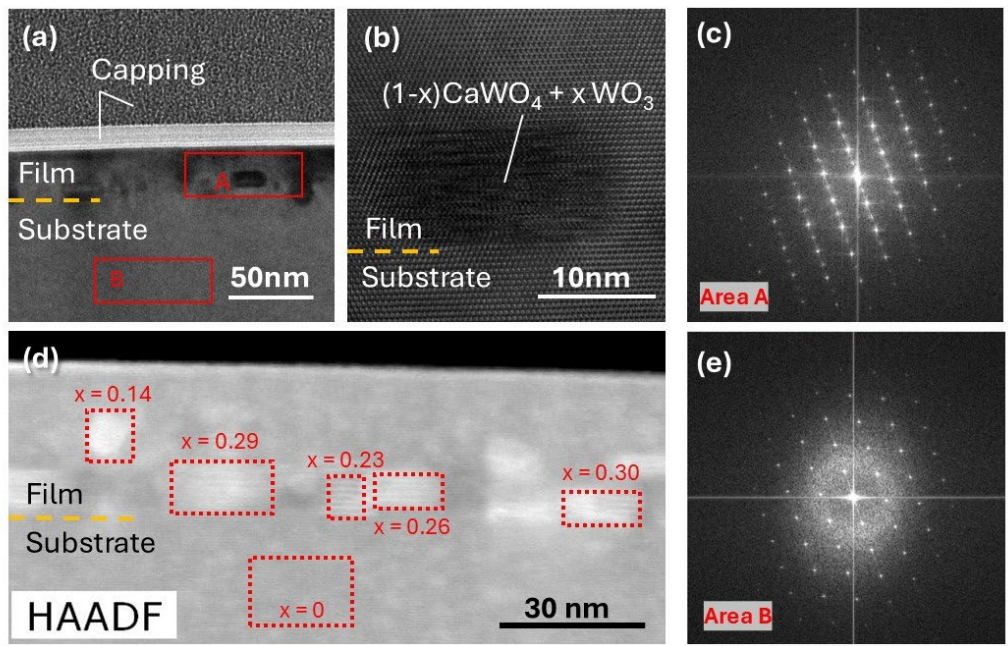

Fig. 7. **Phase separation of annealed, off-stoichiometric CaWO$_4$ thin films.** (a) A high-resolution phase contrast TEM image shows a thin film-substrate interface. (b) High magnification of (a) shows a distorted lattice structure in CaWO$_4$ and WO$_3$ mixed-phase area. (c) Diffraction pattern obtained by Fast-Fourier transform (FFT) of the thin film region shows two sets of lattices. (d) HAADF-STEM image reveals the distribution of Ca and W elements within the thin film region. x, denoting the excess W percentage, is calculated according to EDX quantification. (e) Substrate diffraction pattern obtained by FFT of area B.

We perform TEM measurements on both off-stoichiometric (Fig. 7) and stoichiometric thin films (Fig. 8). In Fig. 7(a) and (b), dark shadowed regions are observed in annealed off-stoichiometric CaWO$_4$ thin films, indicating that the lattice structure is disrupted in these areas. Fast-Fourier transform (FFT) diffraction analysis (Fig. 7(c)) confirms the presence of another phase, which can be seen by comparing the diffraction from the inhomogeneous film with the bare substrate pattern shown in Fig. 7(e). To quantify the stoichiometric ratio, we conduct a STEM-EDX elemental map (see supplement). Fig. 7 (d) shows the distribution of Ca and W elements in these regions, with the value of x denoting excess W ranging from 0.14 to 0.30. The presence of Moiré patterns parallel to the interface in W-excess regions also suggests that there is a different set of lattice planes along $CaWO_4$[001], which is consistent with the observed WO$_3$ (002) peak in the $\theta - 2\theta$ XRD scan (Fig. 3(a)). In contrast, the overall stoichiometric thin film shows a single phase with no other diffraction patterns (Fig. 8(c)) and shows an abrupt interface (Fig. 8(a)) between the substrate and the thin film. In the thin film region, columnar growth distortions are observed starting from the substrate surface. HRTEM (Fig. 8(b) and 8(e)) of the annealed CaWO$_4$ thin film shows a high-quality crystalline structure and no visible film-substrate interface, indicating the thin film possesses the same crystal structure as the substrate. This confirms, from a microscopic perspective, the effectiveness of thermal annealing in removing as-grown defects.

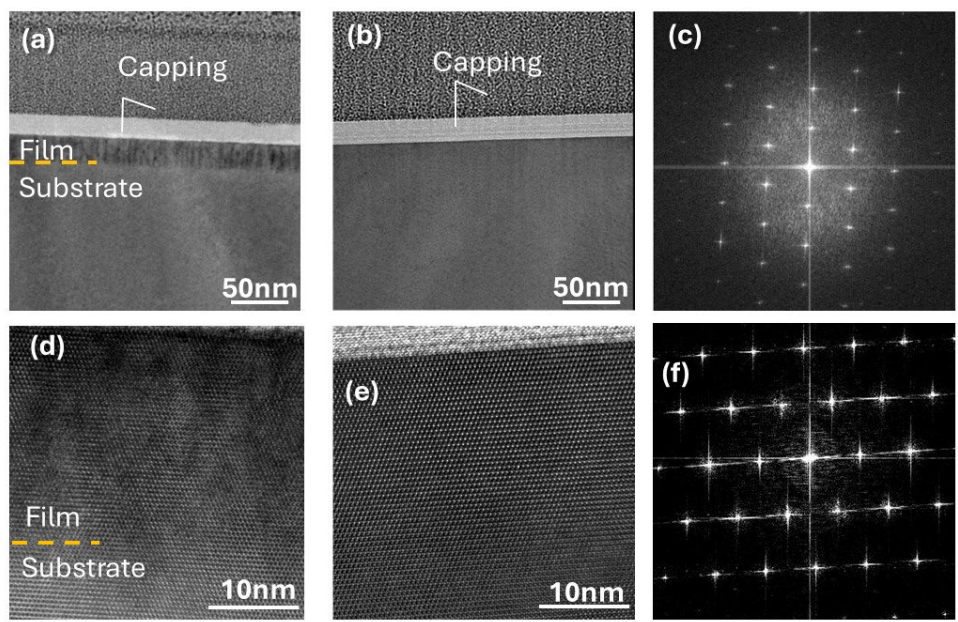

Fig. 8 **Improved microstructure of annealed, stoichiometric thin films.** (a) and (d) TEM images of as-grown stoichiometric CaWO$_4$ thin films at different magnifications. Strips perpendicular to the interface are due to local distortions and defects. (b) and (e) are TEM images of annealed stoichiometric CaWO$_4$ thin films, showing ordered and distortion-free crystalline structures. (c) Diffraction pattern obtained by FFT of (d). (f) Diffraction pattern obtained by FFT of (e).

**Time of flight secondary ion mass spectrometry (TOF-SIMS):**

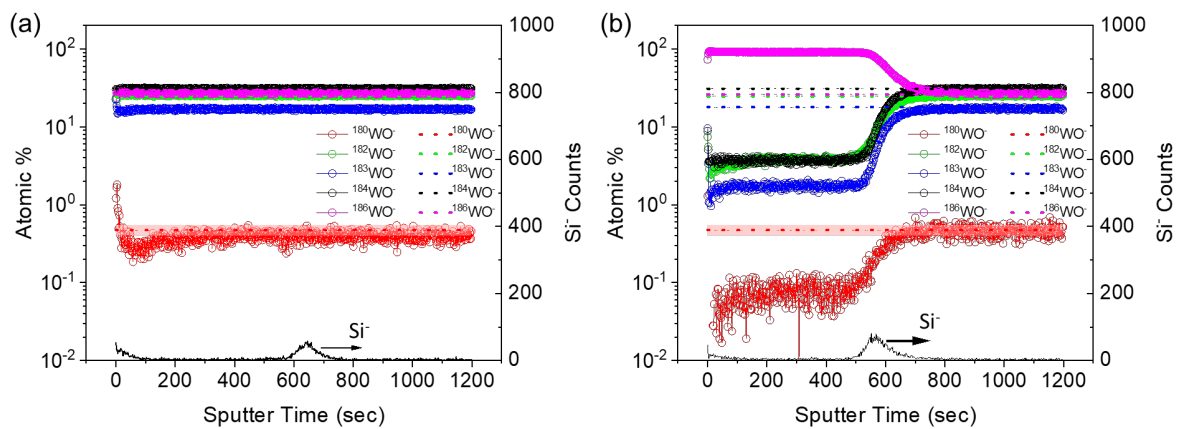

Fig. 9 **Isotopic enrichment**. ToF-SIMS measurement on (a) non-isotope-enriched thin film and (b) isotopically enriched thin film.

ToF-SIMS (Time-of-Flight Secondary Ion Mass Spectrometry) is used to quantify the isotope distribution in the CaWO$_4$ thin film. Fig. 9(a) shows that the isotope abundance in the non-isotope-enriched reference sample is uniformly distributed over the substrate and thin film, which is consistent with the W isotope natural abundance. In the isotopically enriched sample (Fig. 9(b)), there is an abrupt $^{186}$W concentration transition at the thin film-substrate interface (determined by increasing Si$^-$ counts), where the substrate keeps natural abundance while the thin film mainly contains $^{186}$W. More important, the $^{183}$W concentration is only 1. 2% in the thin film, compared with 14.3 % in the substrate. Higher $^{183}$W concentration in MBE-grown thin films than in $^{183}$WO$_3$ source material may be due to residual WO$_3$ with natural abundance in the effusion cell and reused crucible (see supplement). Since $^{183}$W is the only W isotope with non-zero nuclear spin, the average hyperfine coupling of Er$^{3+}$ with the host matrix CaWO$_4$ has been reduced by a factor of 10. As the nuclear spin bath is the major decoherence source at millikelvin temperatures[5], we expect a significant increase of spin coherence time ($T_2$) in the isotopically enriched thin films.

**Spectroscopic Ellipsometry**

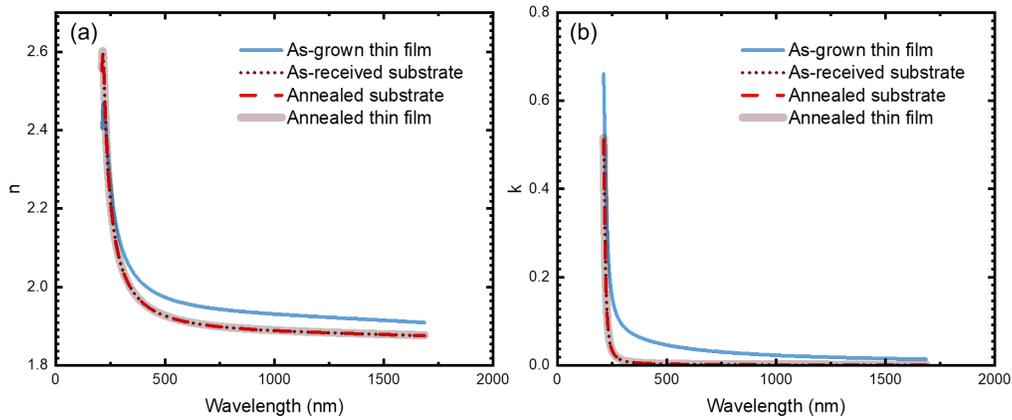

Fig. 10. **Improvement of optical properties upon annealing.** (a) Real part of the refractive index. (b) Imaginary part of the refractive index. Annealing condition is 750°C in pure $O_2$ for 12 hours.

Macroscopic optical properties are important for nanocavity integration applications, since slight changes in refractive index can shift the cavity resonance. To quantify the optical properties of the films grown by MBE, we conducted spectroscopic ellipsometry measurements on the thin films and substrates. Fig. 10 shows the complex refractive index ( $\tilde{n} = n + ik$ ) curve with respect to wavelength. The refractive index dispersion relationship of the as-grown thin films (red) is distinct from the substrates (blue) in which both the real part and the imaginary part of the thin film refractive index are higher than those of the substrates in the telecom-band wavelength regime (1.2-1.6 $\mu m$). We also note that the refractive index of the thin film has a broader resonance peak compared to that of the substrate, indicating a higher scattering rate between the light and defects (e.g., $O_2$ vacancies) in the as-grown thin film[32]. After annealing, the thin film has almost the same refractive index as the substrates, which confirms the elimination of as-grown defects.

**Single Ion Characterization and Optical Spectral Diffusion**

To probe the optical properties of the material at the single-emitter limit, we used an initial off-stoichiometric $CaWO_4$ thin film that was annealed post-growth at 750 °C for 12 hours. $Er^{3+}$ ions were implanted at 35 keV with a low fluence of ($5\times10^9$) ions/cm², targeting a shallow depth compatible with evanescent coupling to surface nanophotonic structures. Following implantation, the sample was annealed at 300 °C for 1 h to partially recover lattice damage induced during the implantation.

In $CaWO_4$, $Er^{3+}$ exhibits a long radiative lifetime (~6.3ms), which makes direct detection and coherent control of a single ion challenging due to low photon collection rates. To increase the emission rate and enable efficient optical addressing, we integrated silicon nanophotonic cavities to provide Purcell enhancement. One-dimensional photonic crystal cavities were fabricated on a separate silicon-on-insulator (SOI) wafer and transferred onto the $CaWO_4$ surface via a flip-chip process[33]. The high-quality factor and small mode volume of these cavities provide strong Purcell enhancement, substantially shortening the effective excited-state lifetime and increasing the detected photon rate. This enhancement enables spectral isolation and measurements of individual $Er^{3+}$ ions. The device assembly was cooled to ~20 mK in a dilution refrigerator with optical access. Photoluminescence excitation (PLE) spectroscopy was performed at zero external magnetic field (|B|=0). In this measurement, the frequency of the pulsed excitation laser was swept across the transition while time-gated, time-delayed fluorescence was detected using superconducting nanowire single-photon detectors (SNSPDs). The fluorescence spectra are measured by scanning the laser frequency in discrete frequency steps using an excitation pulse width of ~5us, followed by a photon-collection window of 180$\mu$s. The measured counts are averaged over $10^4$ repetitions for each discrete laser excitation frequency. Fig. 11 shows a representative zero-field PLE spectrum, where discrete resonance features attributable to individual $Er^{3+}$ ions coupled to the cavity are clearly resolved. The overall spectral spread reflects site-to-site excitation variations of the implanted $Er^{3+}$ ions (introduced by small differences in the local crystal-field environment), which manifest as static, ion-dependent shifts of the observed optical resonances.

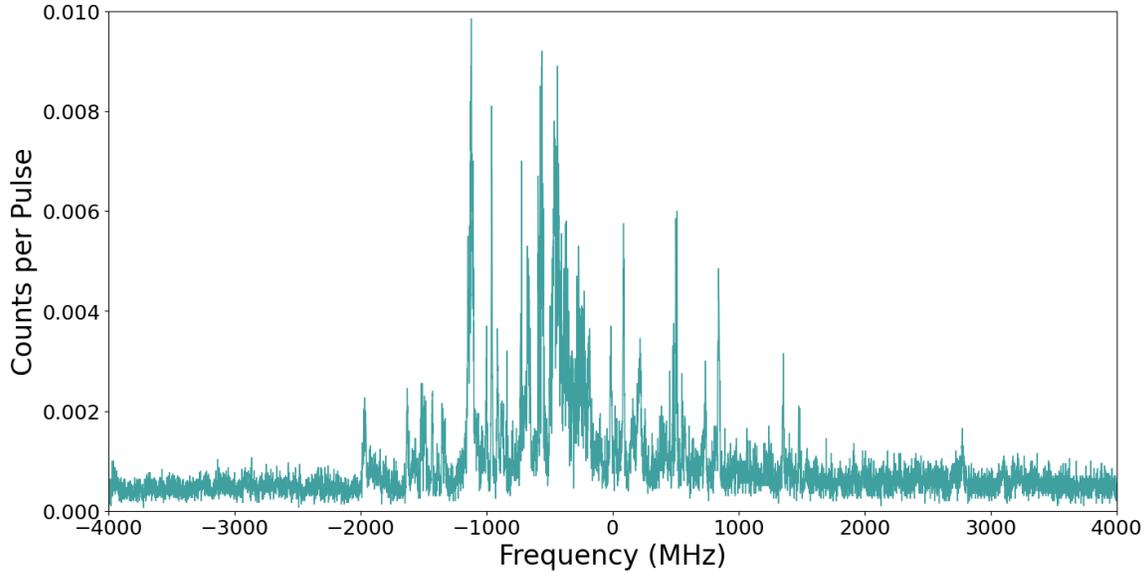

Fig. 11. Zero external field PLE spectrum of $Er^{3+}$ ions coupled to the cavity.

**Optical Spectral Diffusion:**

We emphasize that the following spectral diffusion measurements were performed on an initial off-stoichiometric film and are not representative of the optimized stoichiometric films discussed above. While single ions were successfully resolved, we observed appreciable optical spectral diffusion for ions measured in the MBE-grown epilayer, in comparison to ions implanted into bulk Czochralski-grown $CaWO_4$ crystals processed under comparable conditions. A representative spectral-diffusion trace is shown in Fig. 12(a). The resonance exhibits switching between two discrete frequencies, with each resonance persisting for tens of minutes before transitioning back. Similar two-state switching has been reported for NV⁻ centers in diamond and is likely associated with discrete charge configurations in the local environment.[34,35] In the present system, such switching may arise from different local charge-compensation mechanisms associated with defects or impurities in the bulk or at the surface.[36] We attribute the enhanced spectral diffusion in this initial (off-stoichiometric) epilayer to local strain and compositional inhomogeneities, which act as sites for trapped charges, as hinted in cross-sectional TEM of off-stoichiometric films. To assess the role of post-implant thermal recovery, Fig. 12(b) provides a statistical comparison to bulk $CaWO_4$ reference crystals annealed under two distinct conditions: (i) a standard 1 h furnace anneal at 300 °C and (ii) a rapid thermal anneal (RTA) at 800 °C for 20 s. Although RTA was not part of the primary thin-film processing in this work, the observed linewidth narrowing in the bulk reference suggests that insufficient thermal recovery of implantation damage can contribute to spectral instability. This motivates future exploration of higher-thermal-budget treatments (e.g., RTA) for MBE-grown films, subject to compatibility with film stoichiometry, surface morphology, and device integration. At present, the observed spectral diffusion in the initial off-stoichiometric epilayer complicates long-time coherent protocols (e.g., spin-echo measurements). Ongoing and future work will therefore focus on single-ion PLE and spin-echo measurements in stoichiometric $CaWO_4$ thin films. Consistent with the improved structural quality observed in cross-sectional TEM for stoichiometric films (Fig. 8 (b)), we expect reduced defect-related electric-field fluctuations and correspondingly improved optical stability, approaching that of bulk CZ-grown host crystals.

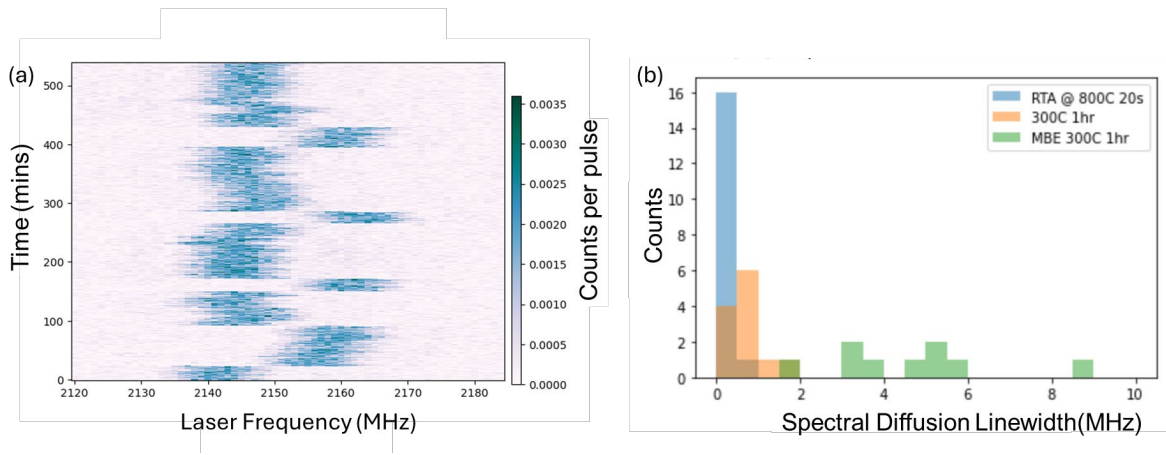

Fig. 12 **Observation of spectral diffusion.** (a) Optical spectral diffusion of a representative $Er^{3+}$ ion implanted in the MBE-grown $CaWO_4$ thin film. (b) Statistical comparison of the spectral diffusion linewidth of different ions measured across the ions implanted in the MBE layer and the bulk CZ-grown host crystal.

**Conclusion**

In summary, we have successfully grown isotopically enriched $Ca^{186}WO_4$ thin films, and ToF-SIMS measurements verify that the $^{183}W$ concentration in the thin films has been reduced from 14.3% to 1.2%. We also demonstrate single-ion addressing in $Ca^{186}WO_4$ thin films and develop methods to characterize macroscopic optical properties of MBE-grown thin films with spectroscopic ellipsometry, as well as techniques to remove as-grown defects via systematic post-annealing treatments. Our work demonstrates a potential platform and process for fabricating highly coherent quantum devices. Future research will focus on spin coherence measurements on thin films, and a substantial increase of the spin coherence time $T_2$ is expected in the isotopically purified $CaWO_4$ thin films.

ACKNOWLEDGMENTS

This work was supported by National Quantum Information Science Research Centers, Co-design Center for Quantum Advantage (C2QA) under Contract No. DE-SC0012704. ToF-SIMS analysis performed by J. D. was supported by the PNNL-OSU Distinguished Graduate Research Program (DGRP) sponsored by U.S. DOE BES Materials Science and Engineering Division under Award 10122 and Oregon State University. A portion of this research was performed on a project award (*10.46936/cpcy.proj.2023.60681/60008771*) from the Environmental Molecular Sciences Laboratory, a DOE Office of Science User Facility sponsored by the Biological and Environmental Research program under Contract No. DE-AC05-76RL01830. This work was performed, in part, at the Center for Integrated Nanotechnologies, an Office of Science User Facility operated for the U.S. Department of Energy (DOE) Office of Science. Sandia National Laboratories is a multimission laboratory managed and operated by National Technology & Engineering Solutions of Sandia, LLC, a wholly owned subsidiary of Honeywell International, Inc., for the U.S. DOE's National Nuclear Security Administration under contract DE-NA-0003525. The views expressed in the article do not necessarily represent the views of the U.S. DOE or the United States Government.

AUTHOR CONTRIBUTION

J.D.T., C.H.A., and F.J.W. conceived the experiments, H.T., K.S., and F.J.W. designed the thin film growth experiments. H.T. and K.S. performed thin film growth. H.T. conducted XRD, AFM, and Ellipsometry measurements. S.P did the TEM proposed by A.K.B. J.A.D carried out ToF-SIMS and analyzed the data with the help of Z.Z. S.P.H, A.T, A.K.B and J.A. fabricated photonic devices and measured single ions PLE data. H.T. J.A.D and A.K.B wrote the manuscript with input from all the authors. F.J.W, Y.G.D, J.D.T., and C.H.A. supervised the project.

COMPETING INTERESTS

The authors declare no competing interests.

Data availability

The data that support the findings of this study are available from the corresponding author upon reasonable request.

Reference

1. Simon, C. *et al.* Quantum memories. *Eur. Phys. J. D* **58**, 1–22 (2010).

2. Kindem, J. M. *et al.* Control and single-shot readout of an ion embedded in a nanophotonic cavity. *Nature* **580**, 201–204 (2020).

3. Awschalom, D. D., Hanson, R., Wrachtrup, J. & Zhou, B. B. Quantum technologies with optically interfaced solid-state spins. *Nat. Photonics* **12**, 516–527 (2018).

4. Ourari, S. *et al.* Indistinguishable telecom band photons from a single Er ion in the solid state. *Nature* **620**, 977–981 (2023).

5. Le Dantec, M. *et al.* Twenty-three–millisecond electron spin coherence of erbium ions in a natural-abundance crystal. *Sci. Adv.* **7**, eabj9786 (2021).

6. Raha, M. *et al.* Optical quantum nondemolition measurement of a single rare earth ion qubit. *Nat. Commun.* **11**, 1605 (2020).

7. Wolfowicz, G. *et al.* Quantum guidelines for solid-state spin defects. *Nat. Rev. Mater.* **6**, 906–925 (2021).

8. Weber, J. R. *et al.* Quantum computing with defects. *Proc. Natl. Acad. Sci. U. S. A.* **107**, 8513–8518 (2010).

9. Alkauskas, A., Bassett, L. C., Exarhos, A. L. & Fu, K.-M. C. Defects by design: Quantum nanophotonics in emerging materials. *Nanophotonics* **8**, 1863–1865 (2019).

10. Stevenson, P. *et al.* Erbium-implanted materials for quantum communication applications. *Phys. Rev. B* **105**, 224106 (2022).

11. Awschalom, D. *et al.* Development of Quantum Interconnects (QuICs) for Next-Generation Information Technologies. *PRX Quantum* **2**, 017002 (2021).

12. Sangouard, N., Simon, C., de Riedmatten, H. & Gisin, N. Quantum repeaters based on atomic ensembles and linear optics. *Rev. Mod. Phys.* **83**, 33–80 (2011).

13. Sipahigil, A. *et al.* An integrated diamond nanophotonics platform for quantum-optical networks. *Science* **354**, 847–850 (2016).

14. Uysal, M. T. *et al.* Spin-Photon Entanglement of a Single $\mathrm{Er}^{3+}$ Ion in the Telecom Band. *Phys. Rev. X* **15**, 011071 (2025).

15. Shin, K. *et al.* Er-doped anatase $TiO_2$ thin films on $LaAlO_3$ (001) for quantum interconnects (QuICs). *Appl. Phys. Lett.* **121**, 081902 (2022).

16. Phenicie, C. M. *et al.* Narrow Optical Line Widths in Erbium Implanted in $TiO_2$. *Nano Lett.* **19**, 8928–8933 (2019).

17. Ji, C. *et al.* Nanocavity-Mediated Purcell Enhancement of Er in $TiO_2$ Thin Films Grown via Atomic Layer Deposition. *ACS Nano* **18**, 9929–9941 (2024).

18. Xie, T. *et al.* Characterization of $Er^{3+}:YVO_4$ for microwave to optical transduction. *Phys. Rev. B* **104**, 054111 (2021).

19. Rančić, M. *et al.* Electron-spin spectral diffusion in an erbium doped crystal at millikelvin temperatures. *Phys. Rev. B* **106**, 144412 (2022).

20. Xu, H. *et al.* Coherent control of interacting solid-state spins below the diffraction limit. Preprint at https://doi.org/10.48550/arXiv.2508.09122 (2025).

21. Masiulionis, I. *et al.* Microstructural and preliminary optical and microwave characterization of erbium doped $CaMoO_4$ thin films. Preprint at https://doi.org/10.48550/arXiv.2508.15122 (2025).

22. Tiranov, A. *et al.* Sub-second spin and lifetime-limited optical coherences in $^{171}Yb^{3+}$:$CaWO_4$. Preprint at https://doi.org/10.48550/arXiv.2504.01592 (2025).

23. Tyryshkin, A. M. *et al.* Electron spin coherence exceeding seconds in high-purity silicon. *Nat. Mater.* **11**, 143–147 (2012).

24. O'Sullivan, J. *et al.* Individual solid-state nuclear spin qubits with coherence exceeding seconds. *Nat. Phys.* **21**, 1794–1800 (2025).

25. Marcks, J. C. *et al.* Nuclear spin engineering for quantum information science. *J. Mater. Res.* **40**, 1433–1448 (2025).

26. Witzel, W. M., Carroll, M. S., Morello, A., Cywiński, Ł. & Das Sarma, S. Electron Spin Decoherence in Isotope-Enriched Silicon. *Phys. Rev. Lett.* **105**, 187602 (2010).

27. Zhang, J. *et al.* Optical and spin coherence of Er spin qubits in epitaxial cerium dioxide on silicon. *Npj Quantum Inf.* **10**, 1–9 (2024).

28. Grant, G. D. *et al.* Optical and microstructural characterization of $Er^{3+}$ doped epitaxial cerium oxide on silicon. Preprint at https://doi.org/10.48550/arXiv.2309.16644 (2023).

29. Seth, S. K. *et al.* Spin Decoherence Dynamics of $\mathrm{Er}^{3+}$ in $\mathrm{CeO}_{2}$ Films. *Phys. Rev. Lett.* **135**, 266901 (2025).

30. Kanai, S. *et al.* Generalized scaling of spin qubit coherence in over 12,000 host materials. *Proc. Natl. Acad. Sci. U. S. A.* **119**, e2121808119 (2022).

31. Tang, H. *et al.* Homoepitaxial growth of CaWO4. *J. Vac. Sci. Technol. A* **42**, 022701 (2024).

32. Fujiwara, H. *Spectroscopic Ellipsometry: Principles and Applications*. (John Wiley & Sons, 2007).

33. Dibos, A. M., Raha, M., Phenicie, C. M. & Thompson, J. D. Atomic Source of Single Photons in the Telecom Band. *Phys. Rev. Lett.* **120**, 243601 (2018).

34. Bluvstein, D., Zhang, Z. & Jayich, A. C. B. Identifying and Mitigating Charge Instabilities in Shallow Diamond Nitrogen-Vacancy Centers. *Phys. Rev. Lett.* **122**, 076101 (2019).

35. Billaud, E. *et al.* Electron paramagnetic resonance spectroscopy of a scheelite crystal using microwave-photon counting. *Phys. Rev. Res.* **7**, 013011 (2025).

36. Becker, F. *et al.* Spectroscopic investigations of multiple environments in Er: CaWO 4 through charge imbalance. *Phys. Rev. Mater.* **9**, 076203 (2025).